\documentclass[a4paper]{jpconf}
\usepackage{graphicx}
\begin{document}
\title{Carbon nanotubes as anisotropic target for dark matter }

\author{G Cavoto$^1$, M G Betti$^1$, C Mariani$^1$, F Pandolfi$^1$, A D Polosa$^1$, I Rago$^1$ and A Ruocco$^2$}

\address{$^1$ Dipartimento di Fisica  and INFN Sapienza Univ. Roma, Piazzale A.Moro, 2 00185 Roma, Italy}

\address{$^2$ Dipartimento di Scienze and INFN, Università degli Studi Roma Tre, Via della Vasca Navale 84, I-00146 Rome, Italy}

\ead{gianluca.cavoto@roma1.infn.it}

\begin{abstract}
Directional detection of Dark Matter (DM)  particles could be accomplished by studying either ion or electron recoils in large arrays of parallel carbon nanotubes. For instance, a MeV mass DM particle could scatter off a lattice electron, resulting in the transfer of sufficient energy to eject the electron from the nanotube surface. The electron can eventually be detected whenever an external electric field is added to drive it from the open ends of the array. This detection scheme would offer an anisotropic response and could be used to select an orientation of the target with respect to the DM wind. A compact sensor, in which the cathode element is substituted with a dense array of parallel carbon nanotubes, could serve as the basic detection unit which - if adequately replicated - would allow to explore a significant region of light DM mass and cross-section. A similar detection scheme could be used to detect DM particles with mass in the GeV range scattering off the surface of a CNT and ejecting a carbon ion. We report about the Monte Carlo simulations of such a system and the R\&D towards a  detector prototype.
\end{abstract}

\section{Introduction }

The particle nature of dark matter (DM)  is still unknown. A vast experimental effort in detecting DM particles present in our Galaxy has been carried out in the last years. It  has been focused on the  investigation  of  a range of masses for the DM particles from tens to several hundreds of GeV or more, motivated by   the  Weakly Interactive Massive Particle paradigm. This program has not yet shown any uncontroversial signal for DM and this calls for extending the search to other mass ranges. In particular, the search  for  DM   with a mass between few  MeV to few  GeV ({\it light} DM)  has also been the subject of  a theoretical  interest \cite{Petraki}\cite{Feng}\cite{Hochberg}\cite{Knapen}.   
	Under the very simple assumption that the galactic  DM particles elastically scatter with ordinary matter,  light targets must be chosen for  favourable kinematics. Carbon nuclei -  for instance  - or     electrons can recoil with a suitable energy to make them detectable. We argue that a carbon target made of  vertically aligned carbon nanotubes (VA-CNTs) can act as a  directional  filter given their anisotropic nature  (Fig.\ref{Filter}) for both  ions or electrons recoiling after the scattering with DM.

   \section{Carbon nanotube as target for dark matter}
   
\begin{figure}[h]
\begin{minipage}{18pc}
\includegraphics[width=18pc]{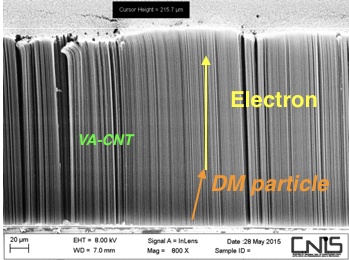}
\caption{\label{Filter} SEM image of a vertically aligned multi-wall  VA-CNTs grown a silicon substrate with the sketch of the idea of electron filtering.}
\end{minipage}\hspace{2pc}%
\begin{minipage}{18pc}
\includegraphics[width=18pc]{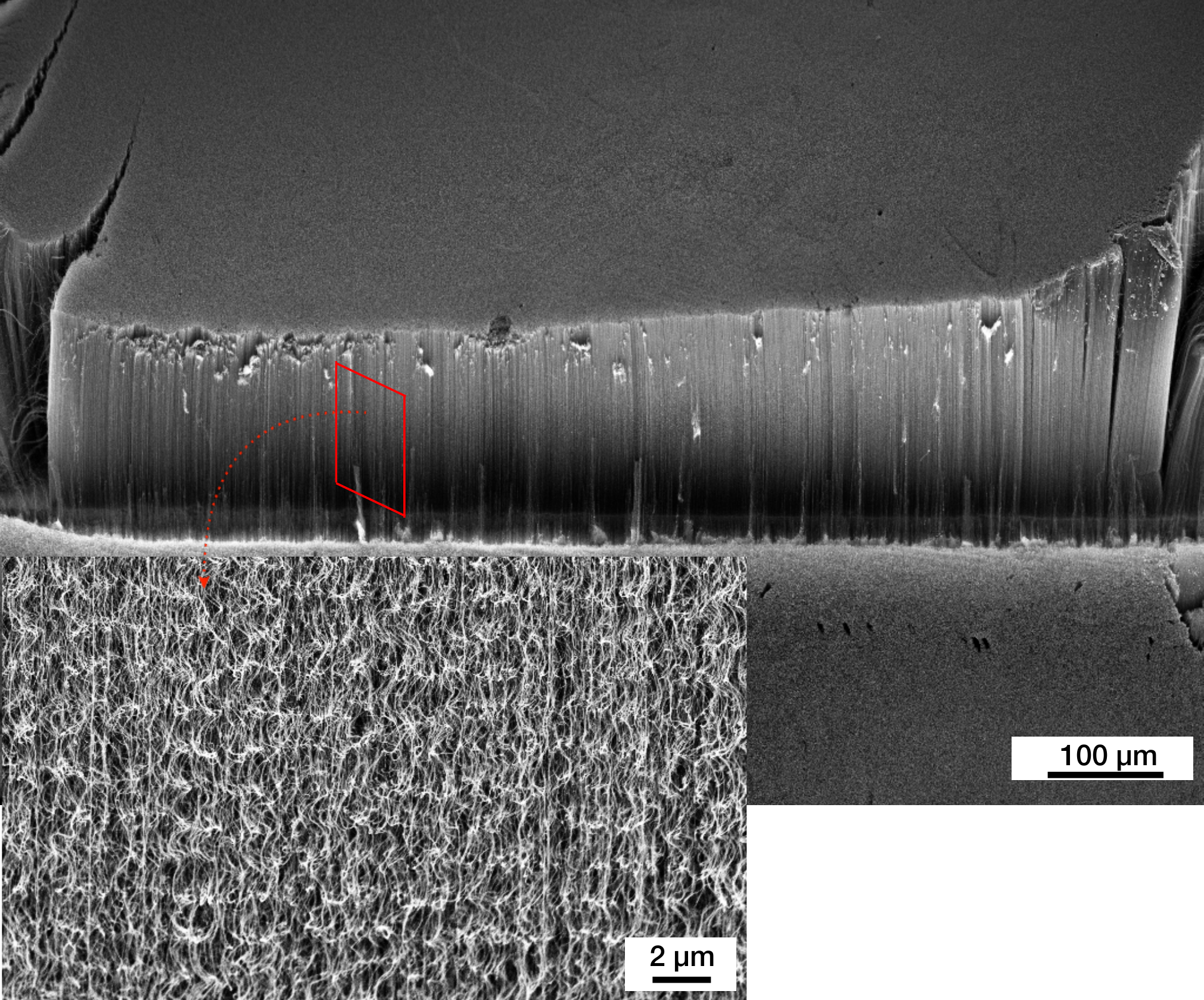}
\caption{\label{SEMCNT2}SEM image of VA-CNTs with an inset showing a detail at smaller scale of the forest.}
\end{minipage} 
\end{figure}

A  CNT   is an intriguing nano-structure  that can be described as a cylinder whose surface is a mono-atomic layer of  carbon atoms - in other words it can be thought as a rolled up graphene sheet \cite{Iijima}. It features an impressive aspect ratio - being its radius 10000 times smaller than its height - and has electric  properties  which depend critically on its chirality.  A single CNT is usually referred to as a single-wall CNT while  a  multi-wall CNT  presents  several coaxial CNTs  grown together. Multi-wall CNTs have a metallic behaviour being therefore good conductors. 
	 VA-CNTs can be synthesized in "forests"  over a silicon or stainless steel substrate  with a chemical vapor deposition technique. Initially,   metal nano-particles are deposited  on the substrate and  then they work as catalysts   by decomposing the  $C_2H_2$  precursor gas injected in an ultra-high vacuum chamber (Fig. \ref{SEMCNT2}) \cite{Rago}.
 
  VA-CNTs have the attractive feature of being hollow structures with atoms confined on the their surfaces. They can be grown to be even 200 $\mu$m tall with a 10 nm diameter and with an inter-spacing that can be engineered to be about 50 nm.  A surface of few 10 cm$^2$ can host about a 10 mg mass available for the DM scattering on the surface of the VA-CNTs.
  In \cite{capparelli} and \cite{Cavoto1} it has been argued that the CNT surface can act as a repulsive barrier to make the carbon ion  bounce within the CNT forest after the DM scattering. This phenomenon would be active for  angles of the carbon ions with respect to the CNT axis up to to a critical angle of about 35$^o$, trapping the ions in the transverse plane. The carbon ion would then exit the CNT forest by moving along the longitudinal direction. Whenever the carbon ion is emitted with a direction that is outside the range allowed by the critical angle, the carbon ions are penetrating the  CNTs and are then quickly stopped.

   \section{Anisotropy of VA-CNTs }
%

 In order to study this anisotropic behaviour, condensed matter techniques have been employed. X-ray photoelectron spectroscopy (XPS) and Raman scattering have been used to investigate  lattice defects induced  on the CNT structure by a very intense  beam  of 5 keV   Ar$^+$  ions  (5 10$^{17}$  ion/cm$^2$). The XPS C-1s line-shape, among other bombardment-induced effects, revealed the spectral evidence of defects associated to C-vacancies \cite{Dacunto}. In the Raman scattering analysis, light can be focused at various heights of the target with a sub-$\mu$m resolution. Also, it can be focused in the interior of the VA-CNT forest, up to a depth 15-20 $\mu$m, that corresponds to several hundred  CNT layers. When the CNT forest is bombarded along a direction parallel to the CNT axis the Raman scattering revealed the presence of defects at all heights. On the contrary, when  it is bombarded along a direction perpendicular to the CNT axis, no defects appear at a depth of 15 $\mu$m.
 More experiments are being conducted to fully explore the angular dependence of the CNT anisotropy to keV  charged ions, possibly validating the model of the charged ion being channeled among the CNT.  If the ions emerging from the CNT are being detected, a light DM detector with a sensitivity to  the DM  direction could be conceived being mostly sensitive DM particles with mass of a few GeV.

    \section{Detecting electrons, a dark-PMT }
    
       DM particles could be as light as few MeV. In this case, assuming a non-zero coupling of DM to electrons, a detection scheme based on DM-electron scattering would be very appealing.  
       
               \begin{figure}[h]
\begin{center}
\includegraphics[width=18pc]{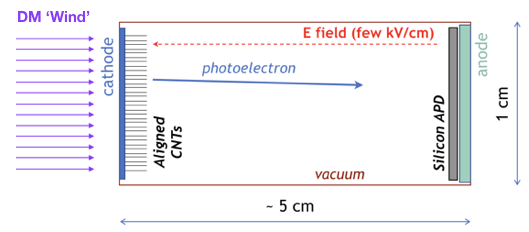}
\caption{\label{darkPMT} Conceptual design of the {\it dark-PMT} unit.}
\end{center}
\end{figure}

 A conceptual scheme (a {\it dark-PMT} )  for this is shown in Fig. \ref{darkPMT}. The VA-CNTs are arranged to be a cathode layer placed a in vacuum tube. An electric field of a few kV/cm is applied to accelerate an electron emerging from the "dark" photocathode and to guide it  to the anode. The anode can host an active silicon detector (e.g. APD or SDD) to measure each single electron.
  Since the work function of a CNT for the electron photo-emission  is relatively  high ( $>$ 4.7 eV)   \cite{Peng}  such device would be blind to visible light and could operate even at room temperature being  thermal photo-emission   suppressed as well. However, the  electric field applied between the anode and the cathode should be not too high  to prevent  field emission from the CNTs themselves but large enough to make the electrons detectable at the anode (more details in \cite{Cavoto2}).
 
An electron scattered by  MeV DM particles would have a few eV kinetic energy. 
While data on electron-CNT interaction at this energy are not available, in  \cite{Cavoto2} it is argued that the dark photocathode would still feature an anistropic behaviour, acting as a filter for electrons emitted in a direction collinear to the CNT axis. By arranging few thousand  units of dark PMTs with a total  exposure of 1 kg $\cdot$ y,  a sensitivity to a DM-electron cross section  as low as 10$^{-40}$ cm$^{2}$ for a 10 MeV DM particle can be reached.

\begin{figure}[h]
\begin{minipage}{18pc}
\includegraphics[width=18pc]{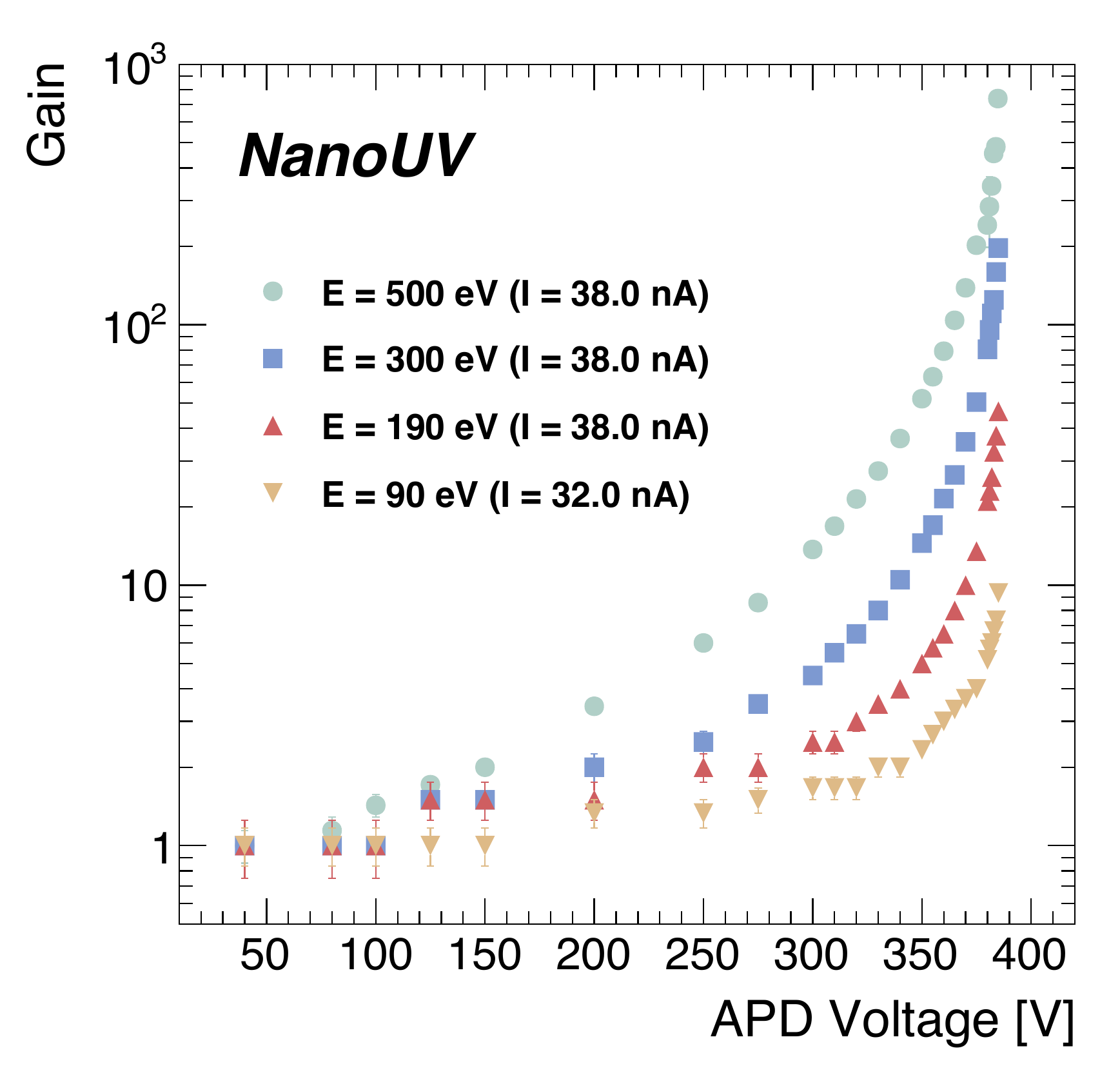}
\caption{\label{GainAPD}  Normalized APD current  (to current with $V_{APD}$ = 40 $V$)  for different  energy of the electron beam as a function of $V_{APD}$.}
\end{minipage}\hspace{2pc}%
\begin{minipage}{18pc}
\includegraphics[width=18pc]{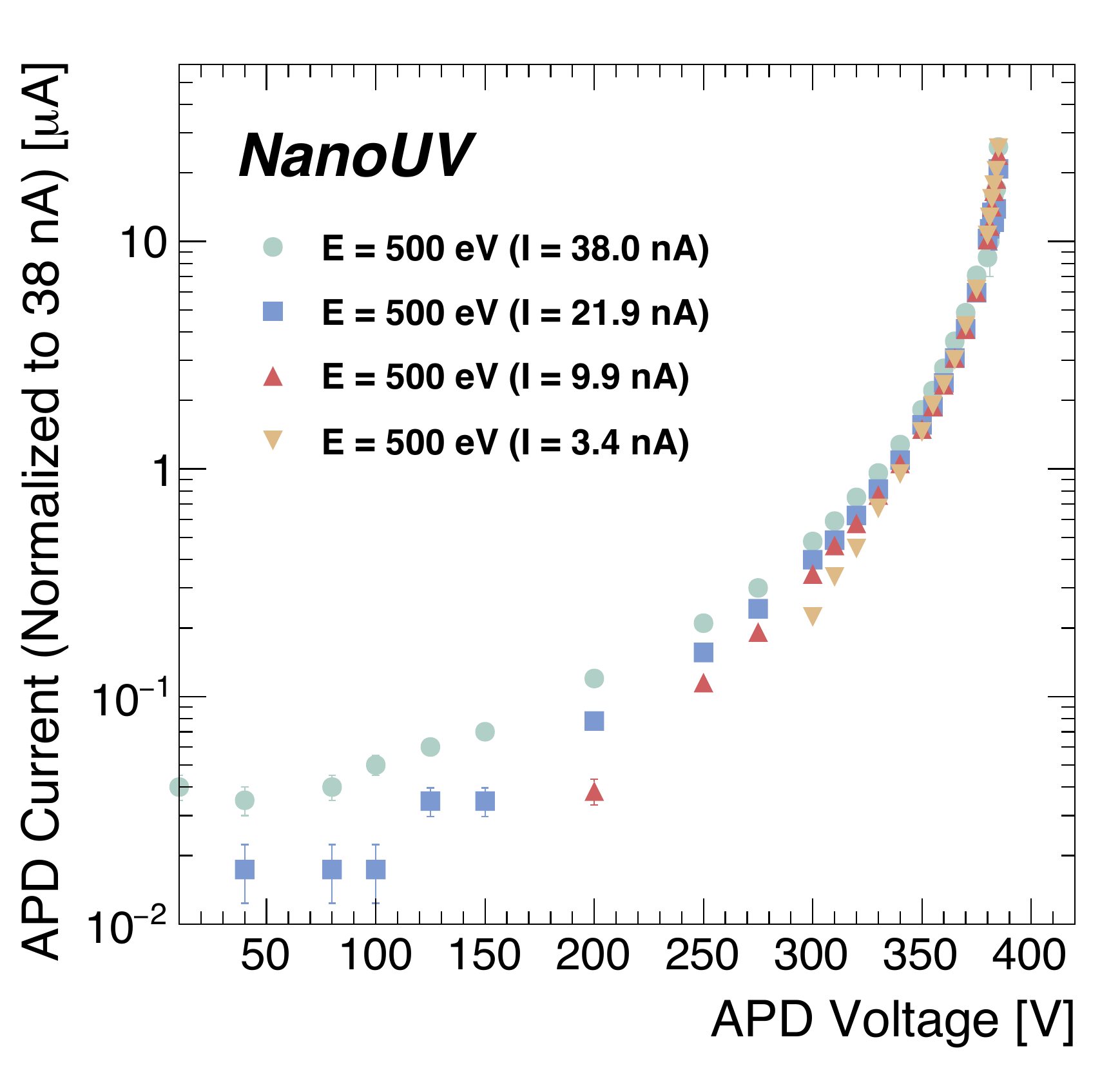}
\caption{\label{APDcurrent} Normalized APD current  (to current with $V_{APD}$ = 40 $V$)  for different electron beam current ($E$ = 500 eV) as a function of $V_{APD}$.}
\end{minipage} 
\end{figure}

A prototype dark PMT is currently being assembled with some of its  key elements already being tested. VA-CNTs  have been grown up to the record height of 170 $\mu$m   on a substrate transparent to UV-light  (fused silica),  eventually allowing a  test of photo-emission with a  UV radiation able to  overcome the 4.7 eV work function barrier. 
A special APD with no protective layer and a very thin  Si oxide cover is considered for the dark PMT.  A demonstration of this APD being able to  detect electrons with an energy ranging from 90 eV to 500 eV  (Fig.2) has been obtained. This was done by using an electron beam in an ultra-high vacuum facility with an electron gun with few tens  meV energy resolution and  per-mille  current control (Fig. \ref{GainAPD} and \ref{APDcurrent}).

    \section{Conclusions }
    The quest for the DM particle nature is still open and new experimental techniques must be explored to be sensitive to faint and rare experimental signatures. The use of carbon nanostructures (see also \cite{hochberg2}) can open a new window of sensitivity to the sub-GeV  DM mass range. VA-CNTs might allow to have a suitable target mass in a limited volume and might have the additional feature of being sensitive to the DM direction, an additional signature to suppress background events in the search for DM.    A prototype detector  (a dark PMT) is being assembled with a calibration run with UV-light foreseen in the near future.

   \section{Acknowledgments }
   We are grateful to our colleagues  L.Casalis and A.Goldoni (Elettra Sincrotrone Trieste) for the use of their CVD facility to grow our VA-CNTs. We acknowledge the support of   EU ATTRACT  (NanoUV project),  Sapienza  Grande Progetto di Ateneo 2017,  Amaldi  ResearchCenter (MIUR program ”Dipartimento diEccellenza” CUP:B81I18001170001) and INFN CNS2.

\end{document}